\documentclass[floats, prd, eqnum, showpacs, nofootinbib, 
twocolumn, 
dvipdfmx, 
eqsecnum]{revtex4-1}

\usepackage{color,graphicx}
\usepackage{amsfonts}
\usepackage{amssymb}
\usepackage{comment}
\usepackage{ulem}
\usepackage{hyperref}
\hypersetup{
 setpagesize=false,
 bookmarksnumbered=true,%
 bookmarksopen=true,%
 colorlinks=true,%
}

\begin{document}

\title{Constraints on the black-hole charges of M87* and Sagittarius~A* by changing rates of photon spheres can be relaxed}
\author{Naoki Tsukamoto}\email{tsukamoto@rikkyo.ac.jp}
\author{Ryotaro Kase}

\affiliation{
Department of Physics, Faculty of Science, Tokyo University of Science, 1-3, Kagurazaka, Shinjuku-ku, Tokyo 162-8601, Japan }

\begin{abstract}
The Event Horizon Telescope (EHT) Collaboration observed ring images called the shadows of M87* and Sagittarius~A* (Sgr~A*), which are supermassive objects in M87 and our galaxy, respectively, and their general relativistic magnetohydrodynamic simulations of black holes imply that the observed rings are formed by the gravitational lensing of synchrotron radiations from a hot plasma near outside of supermassive black holes. The EHT Collaboration gave constrains on the electrical or alternative charges of M87* and Sgr A* under an assumption that the radius of the observed ring should be proportional to the changing rates of photon spheres by the charges. Since the validness of this assumption is not sure, it is worth to checking the same constraints under another assumption. In this paper, we consider the changing rates of not only the photon spheres but also lensing rings in a simple model and we test whether aforementioned constraint is robust. We conclude that EHT Collaboration's constraints based on the changing rates of the photon spheres can be relaxed compared to that based on the changing rate of the lensing rings while we do not claim that the observed rings are formed by the photon spheres and the lensing rings in our simple model. We concentrate on Reissner-Nordstr\"{o}m black hole spacetimes in this paper, but our result implies the relaxation of the bound of the charge parameters on other black hole spacetimes.
\end{abstract}

\date{\today}

\maketitle

\section{Introduction}
Black holes predicted in general relativity had been considered as a hypothetical compact object for a long time, but, recently, their astrophysical roles in universe are seriously considered since the direct detection of gravitational waves emitted from stellar-mass black holes was reported by LIGO Scientific Collaboration and Virgo Collaboration~\cite{Abbott:2016blz}.     
It is known that the black holes or alternative compact objects with a strong gravitational field  
have the unstable circular orbits of light rays which form a so-called photon sphere~\cite{Claudel:2000yi,Perlick_2004_Living_Rev}. 
Rays emitted by a disk can wind around the photon sphere. 
We consider rays that are emitted from the disk and cross the disk plane $n$ times before reaching an observer.
We can classify ring images of the rays projected on an observer plain by using $n$ 
but we should be careful for various terms used in the literatures~\cite{Perlick:2021aok}.  
On this paper, we use terms a direct ring~($n=0$), a lensing ring~($n=1$), and photon rings~($n \geq 2$) as following Ref.~\cite{Gralla:2019xty}.
In astrophysics, the rings with $n \geq 1$ are often called photon rings~\cite{Johannsen:2010ru}, 
or the rings with $n=1$ and $n \geq 2$ are called that $n=1$ photon ring and higher-order photon rings~\cite{Broderick:2021ohx}, respectively\footnote{
In gravitational lensing in a strong gravitational field, retrolensing images ($n=1$)~\cite{Holz:2002uf,Eiroa:2003jf} and relativistic images ($n=2$)~\cite{Virbhadra_Ellis_2000} are also used.}.    A lensing ring near outside of the photon sphere can be one of the observational evidences of the strong gravitational field 
around the compact objects. The appearances of lensing rings and photon rings
were investigated often~\cite{Hagihara_1931,Darwin_1959,Atkinson_1965,Ames_1968,Synge:1966okc,Luminet_1979,Ohanian_1987,Fukue,Nemiroff_1993,Falcke:1999pj,Frittelli_Kling_Newman_2000,Virbhadra_Ellis_2000,Bozza_Capozziello_Iovane_Scarpetta_2001,Perlick:2003vg,Bozza:2007gt,Hioki:2009na,Bozza_2010,Virbhadra:2022iiy,Tsupko:2022kwi,Soares:2023uup,Ali2024}.

In 1918, Curtis found that the giant elliptical galaxy M87 has a filament~\cite{Curtis_1918} which was called a jet by Baade and Minkowski~\cite{Baade}.
The jet was observed in the radio~\cite{Bolton,Kassim1993,Owen:2000vi}, optical~\cite{Biretta1999,Perlman:2011ba}, x-ray~\cite{Marshall:2001de}, and gamma-ray bands~\cite{HESS:2011huh}. 
In the radio bands, M87 is seen as an extended bright structure believed to be powered by a central supermassive compact object called M87*~\cite{Owen:2000vi}. 
The synchrotron age of the radio halo is estimated as $4 \times 10^7$ years and the jet has a kinetic power of $10^{37}\sim 10^{38}$~J/s~\cite{Owen:2000vi,deGasperin:2012id,Broderick:2015tda}.  
The powerful jet might be powered by rotating energy extracted from a central supermassive black hole in an electromagnetic environment~\cite{Penrose:1969pc,Blandford:1977ds}. 

The supermassive compact objects M87* in the center of M87 and Sagittarius~A* (Sgr~A*) in the center of the Milky Way galaxy
are black hole candidates with the largest apparent size seen from us.  
Recently, the Event Horizon Telescope (EHT) Collaboration reported the ring images of M87*~\cite{EventHorizonTelescope:2019dse,EventHorizonTelescope:2019ggy} 
and Sgr~A*~\cite{EventHorizonTelescope:2022wkp}
by using a global very-long baseline interferometry array observing at a wavelength of $1.3$~mm. 
They confirmed that the ring images are consistent with neither the photon spheres nor the photon rings 
but the strong gravitational lensed images which are formed synchrotron radiations from a hot plasma 
which is near outside of the photon rings of the Kerr black holes in ray-traced general-relativistic magnetohydrodynamic (GRMHD) simulations.
The dominant contribution to the ring image observed by the EHT would be direct emissions ($n=0$) from the light source while a minor contribution from the lensing ring ($n=1$) would exist and they are hardly separable from each other. The photon rings ($n \geq 2$) contribute little to the observed image. We might separate the lensing ring ($n=1$) from the other components and detect it in future space observations~\cite{Lupsasca:2024xhq,Johnson:2024ttr}.

A Reissner-Nordstr\"{o}m black hole with a mass $M$ and an electrical charge $Q$, which is the static, spherically symmetric, asymptotically-flat, electrovacuum solution of Einstein-Maxwell equations,
is often investigated as the second simplest black hole spacetime among black hole spacetimes including the Schwarzschild spacetime as its special case. 
It is difficult to give strict constraints on the charge $Q$  
from observations in a weak gravitational field, such as the time delay of lensed rays~\cite{Sereno:2003nd} and microlensing~\cite{Ebina:2000dg}, 
since the effects of the charge on observables are tiny.
Therefore, we should consider phenomenon in the strong gravitational field to distinguish it from the Schwarzschild black hole.
The shadow image of the Reissner-Nordstr\"{o}m black hole~\cite{deVries:2000,Zakharov:2005ek,Takahashi:2005hy,Zakharov:2014lqa,daSilva:2023jxa} or 
its relativistic images which appear near outside the photon sphere in the context of the gravitational lensing in the strong gravitational field~\cite{Eiroa:2002mk,Bozza:2002zj,Eiroa:2003jf,Bin-Nun:2010exl,Bin-Nun:2010lws,Tsukamoto:2016oca,Tsukamoto:2016jzh,Tsukamoto:2022uoz,Tsukamoto:2022tmm,Sasaki:2023rdf,Aratore:2024bro,Hsieh:2021scb,Hsieh:2021rru} were investigated.
The observables in overcharged cases~\cite{Shaikh:2019itn,Tsukamoto:2021fsz,Tsukamoto:2021lpm,Tsukamoto:2020iez, Chiba:2017nml,Wu:2024ixf} were also studied.

Recently, the EHT Collaboration has constrained the ratios of an electrical charge $Q$ to a mass $M$  
as $0\leq \left|Q \right|/M<0.90$ for M87*~\cite{EventHorizonTelescope:2021dqv} and 
$0 \leq \left|Q \right|/M \leq 0.84$ for Sgr~A*~\cite{EventHorizonTelescope:2022xqj} in the $1\sigma$ level 
by using the deviation of the photon sphere of Reissner-Nordstr\"{o}m black hole from the Schwarzschild black hole.   
There remains a question how accurate this constraint is
since the observed rings are consistent with not the photon sphere
but the gravitational lensed images by synchrotron radiations from a hot plasma~\cite{EventHorizonTelescope:2019dse,EventHorizonTelescope:2019ggy}.

Gralla \textit{et al.} considered the image of the Schwarzschild black hole around an optically and geometrically thin accretion disk~\cite{Gralla:2019xty}. 
They confirmed that the photon ring, which makes the very narrow spike of intensity, is negligible for optically thin emission
and that the lensing ring, which is larger than the photon ring by a few percent, can determine the observed ring size.  
The lensing ring by the static and spherically symmetric black hole surround by the disk model~\cite{Gralla:2019xty} is too simple 
to explain the jet from M87 and the asymmetry of the observed ring images~\cite{EventHorizonTelescope:2019pgp} 
and 
their polarization~\cite{EventHorizonTelescope:2021bee,EventHorizonTelescope:2021srq,EventHorizonTelescope:2023fox,EHT2024a,EHT2024b}. 
The position of the lensing rings do not match observed images 
but the changing rate of the lensing ring is useful to study whether the constraints on the charges by the change rate of photon sphere in Refs.~\cite{EventHorizonTelescope:2021dqv,EventHorizonTelescope:2022xqj} is robust.

On this paper, we show that the constraint of the charge by using the change rate of the photon sphere can be relaxed by comparing with constraint from the change rate of lensing rings in the simple disk and black hole system.
For simplicity, we do not treat the angular momentum of the black holes.
The angular momentum does not contribute the ring size much  
and the shadow observations permit a wide range of the angular momentum of the Kerr black hole 
while nonzero angular momentum is preferable to explain the asymmetry of the ring images  
and the jet emitted from M87*~\cite{EventHorizonTelescope:2019pgp}.

We concentrate on the Reissner-Nordstr\"{o}m black hole around the geometrically thin disk with an inner edge at the innermost stable circular orbit (ISCO) of a particle as the simplest model.   
One may consider that electrical charged black holes in nature would be neutralized soon due to plasma around the black hole. 
We do not care about the neutralization of the charged black hole and the electromagnetic interactions between the charged black hole and its astrophysical environment such as electrons and magnetic fields, 
which are the same assumptions as Refs.~\cite{EventHorizonTelescope:2021dqv,EventHorizonTelescope:2022xqj}.   
This is because our purpose is to construct a method by the lensing ring to give more certain constraints on not only the electrical charge of the black hole but also on the parameters of alternative black hole spacetimes. 
In order to omit on various electromagnetic effects caused by the electrical charge, 
one can read the Reissner-Nordstr\"{o}m spacetime as nonelectrically-charged spacetimes with the same metric~\cite{Bin-Nun:2009hct,Zakharov:2018awx,Babichev:2017guv,Badia:2017art,Zakharov:2021gbg}.

This paper is organized as follows. We review the Reissner-Nordstr\"{o}m black hole spacetime in Sec.~II 
and a setup for imaging in Sec.~III. 
In Sec.~IV, we investigate the lensing ring images of rays emitted from the ISCO. 
In Sec.~V, we review the observations on M87* and Sgr~A* and constrain the charges by using change rate of radius of the lensing ring.
In Sec.~VI, we discuss and conclude our results.
We note the analytic expressions of radii of the ISCO and a periastron in Appendix~A 
and 
we comment on a ring image of M87* at wavelength of $3.5$~mm reported in Ref.~\cite{Lu:2023bbn} in Appendix~B.
We use units in which the light speed and Newton's constant are unity.

\section{Reissner-Nordstr\"{o}m black hole}
A line element in a Reissner-Nordstr\"{o}m spacetime, which is the electrovacuum solution of Einstein-Maxwell equations, is expressed by, in coordinates $x^\mu=(t, r, \vartheta, \varphi)$,
\begin{equation}
\mathrm{d}s^2=-f(r)\mathrm{d}t^2+\frac{\mathrm{d}r^2}{f(r)}+r^2 ( \mathrm{d}\vartheta^2 +\sin^2\vartheta \mathrm{d}\varphi^2),
\end{equation}
where a function $f(r)$ is defined by 
\begin{equation}
f(r) \equiv 1- \frac{2M}{r} + \frac{Q^2}{r^2},
\end{equation}
where $M$ and $Q$ are the mass and electrical charge of a central object, respectively. 
The spacetime has an event horizon at $r=r_\mathrm{H}$, where 
\begin{equation}
r_\mathrm{H}\equiv M +\sqrt{M^2-Q^2}\,,
\label{eq:rH}
\end{equation}
for $\left| Q \right| \leq M$ and it has naked singularity for $\left| Q \right| > M$.
The radius of the event horizon monotonically decreases from $2M$ to $M$ as the charge $\left| Q \right|$ increases from $0$ to $M$.
There are time-translational and axial Killing vectors 
\begin{equation}
t^\mu \partial_\mu = \partial_t\quad {\rm and}\quad \varphi^\mu \partial_\mu=\partial_\varphi\,, 
\label{Killing}
\end{equation}
because of the stationarity and axial symmetry of the spacetime, respectively. 
We set $\vartheta=\pi/2$ without loss of generality. 

The radius of the ISCO of a particle with a mass is given by the largest positive root $r=r_\mathrm{ISCO}$ of an equation 
\begin{equation}
r^3-6Mr^2+9Q^2r-\frac{4Q^4}{M}=0,
\label{eq:rISCO}
\end{equation}
which is shown as Eq.~(100) on page 223 in Ref.~\cite{Chandrasekhar:1985kt}. Its analytic form is shown in Appendix~A.
The ISCO radius $r_\mathrm{ISCO}$ monotonically decreases from $6M$ to $4M$ as the charge $\left| Q \right|$ increases from $0$ to $M$.

The trajectory of a ray is given by 
\begin{equation}\label{eq:trajectory1}
-f(r)\dot{t}^2+\frac{\dot{r}^2}{f(r)}+r^2 \dot{\varphi}^2=0,
\end{equation}
where the dots denote a differentiation with respect to an affine parameter along the ray.
We define the impact parameter of the light ray as $b\equiv L/E$, where 
\begin{equation}\label{eq:defE}
E\equiv -g_{\mu \nu} t^{\mu} \dot{x}^\nu = f(r)\dot{t}>0
\end{equation}
and 
\begin{equation}\label{eq:defL}
L\equiv g_{\mu \nu} \varphi^{\mu} \dot{x}^\nu=r^2\dot{\varphi},
\end{equation}
are the conserved energy and angular momentum of the ray 
associated with the Killing vectors $t^{\mu}$ and $\varphi^{\mu}$ given in Eq.~(\ref{Killing}), respectively.
At the periastron $r=P$ of the ray, 
from Eq.~(\ref{eq:trajectory1}), we get a relation 
\begin{equation}\label{eq:trajectory2}
f_\mathrm{P}\dot{t}_\mathrm{P}^2=P^2 \dot{\varphi}_\mathrm{P}^2,
\end{equation}
where the values at the periastron are denoted by the subscript~P, since 
$\dot{r}_\mathrm{P}$ should vanish. 
We note that the periastron is often called the closest distance of the ray in gravitational lensing.
By using Eq.~(\ref{eq:trajectory2}), we can express the impact parameter as 
\begin{equation}\label{eq:trajectory3}
b=b_\mathrm{P}=\frac{L_\mathrm{P}}{E_\mathrm{P}}= \frac{P^2 \dot{\varphi}_\mathrm{P}}{f_\mathrm{P} \dot{t}_\mathrm{P}}= \pm \frac{P}{\sqrt{f_\mathrm{P}}},
\end{equation}
where the sign $\pm$ should be chosen to be equal with the sign of $b$, $L$, and $\dot{\varphi}$.

The trajectory of the ray~(\ref{eq:trajectory1}) can be expressed by
\begin{equation}\label{eq:trajectory4}
\dot{r}^2+V(r,b)=0,
\end{equation}
where $V(r,b)$ is an effective potential defined by
\begin{equation}\label{eq:effective}
V(r,b)\equiv E^2 \left( f(r) \frac{b^2}{r^2} -1 \right).
\end{equation}
The ray can be in a place for $V(r,b)\leq 0$. From $V(\infty,b)= -E^2 \leq 0$, the ray can be at a spatial infinity. 
Rays, which fall to a black hole from the spatial infinity, reach to the event horizon if their impact parameter is $\left|b \right| < b_\mathrm{ph}$, 
where $b_\mathrm{ph}$ is a critical impact parameter defined by
\begin{equation}\label{eq:trajectory5}
b_\mathrm{ph}\equiv \frac{r_\mathrm{ph}}{\sqrt{f(r_\mathrm{ph})}},
\end{equation}
where $r_\mathrm{ph}$ is the radius of a unstable circular light orbit given by
\begin{equation}\label{eq:rph}
r_\mathrm{ph}\equiv \frac{3M+\sqrt{9 M^2-8 Q^2}}{2},
\end{equation}
otherwise they are scattered at their periastrons and return to the spatial infinity.
We can confirm 
\begin{equation}
V(r_\mathrm{ph},b_\mathrm{ph})=\frac{\partial}{\partial r} V(r_\mathrm{ph},b_\mathrm{ph})=0
\end{equation}
and 
\begin{equation}
\frac{\partial^2}{\partial r^2} V(r_\mathrm{ph},b_\mathrm{ph})<0
\end{equation}
in a straightforward calculation.
The radius of the unstable circular light orbit $r_\mathrm{ph}$  
monotonically decreases from $3M$ to $2M$ as the charge $\left| Q \right|$ increases from $0$ to $M$.
Figure~1 shows the specific radii $r_\mathrm{H}/M$, $r_\mathrm{ph}/M$, and $r_\mathrm{ISCO}/M$ as the functions of $Q/M$. 
\begin{figure}[htbp]
\begin{center}
\includegraphics[width=\linewidth]{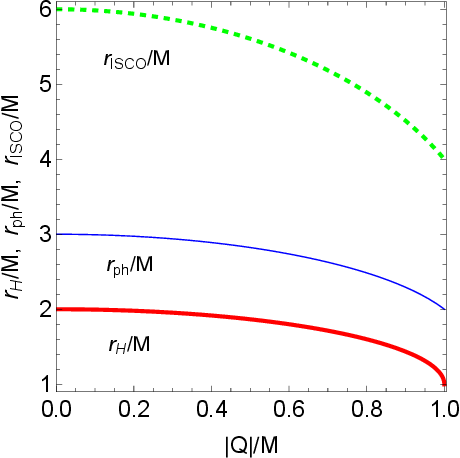}
\end{center}
\caption{
The specific radii $r_\mathrm{H}/M$, $r_\mathrm{ISCO}/M$, and $r_\mathrm{ph}/M$ 
given by Eqs.~(\ref{eq:rH}), (\ref{eq:rISCO}), and (\ref{eq:rph}), respectively,
against the specific charge $\left| Q \right|/M$ are shown. 
Thick-Dashed~(green), thin-solid~(blue), and thick-solid~(red) curves denote $r_\mathrm{ISCO}/M$, $r_\mathrm{ph}/M$, and $r_\mathrm{H}/M$, respectively.}
\end{figure}

We can rewrite the trajectory of the ray~(\ref{eq:trajectory1}) in
\begin{equation}\label{eq:trajectory6}
\left( \frac{\mathrm{d} r}{\mathrm{d} \varphi} \right)^2= r^2 \left( \frac{r^2}{b^2}-f(r) \right). 
\end{equation}
By taking the square root of Eq.~(\ref{eq:trajectory6}) and integrating it from an initial radius $r_\mathrm{e}$ at which the ray is emitted to a final radius $r_\mathrm{o} (>r_\mathrm{e})$ 
at which it is observed,
the deflection angle $\gamma$ of the ray, which sweeps during its travel, 
is expressed by
\begin{equation}\label{eq:deflection_angle0}
\gamma = 
\pm \int^{r_\mathrm{o}}_{r_\mathrm{e}} \frac{\mathrm{d}r}{r\sqrt{{r^2}/{b^2}-f(r)}},
\end{equation}
where we should choose the sign of $\pm$ to be the same as the sign of ${\mathrm{d} r}/{\mathrm{d} \varphi}$, 
and where the deflection angle of the ray is defined by 
\begin{equation}\label{eq:deflection_angle2}
\gamma \equiv  \int^{\varphi_\mathrm{o}}_{\varphi_\mathrm{e}} \mathrm{d}\varphi = \varphi_\mathrm{o}-\varphi_\mathrm{e}. 
\end{equation}
Hereinafter, we set the azimuthal angle coordinate $\varphi$ to be a final azimuthal angle $\varphi_\mathrm{o}=\pi/2$. 
An initial azimuthal angle $\varphi_\mathrm{e}$ and the deflection angle $\gamma$ take values in ranges $-\infty < \varphi_\mathrm{e} < \infty$ and $-\infty < \gamma < \infty$, respectively.
Note that ${\mathrm{d} r}/{\mathrm{d} \varphi}$ changes its sign only when the ray passes the periastron at $r=P=P(b)$.
If it passes the periastron during the travel,
we calculate the deflection angle by   
\begin{equation}\label{eq:deflection_angle3}
\gamma = - \int^{P(b)}_{r_\mathrm{e}} \frac{b \mathrm{d}r}{r\sqrt{r^2-f(r)b^2}}
+ \int^{r_\mathrm{o}}_{P(b)} \frac{b \mathrm{d}r}{r\sqrt{r^2-f(r)b^2}},
\end{equation}
and, if else, by
\begin{equation}\label{eq:deflection_angle1}
\gamma = \int^{r_\mathrm{o}}_{r_\mathrm{e}} \frac{b \mathrm{d}r}{r\sqrt{r^2-f(r)b^2}}. 
\end{equation}
The analytic form of the periastron $P(b)$ is shown in Appendix~A.

\begin{figure*}[t]
\includegraphics[width=0.75\linewidth]{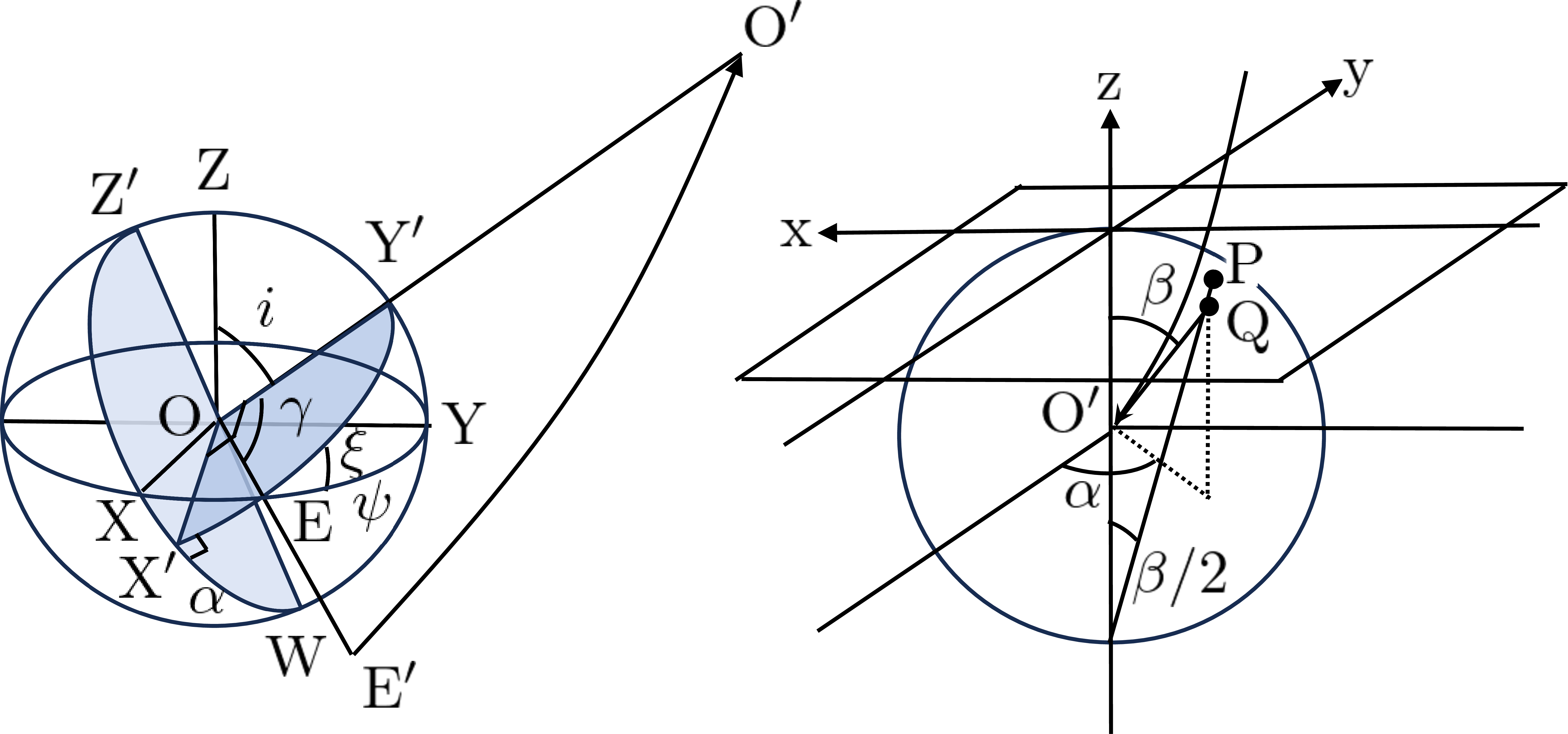}
\caption{
\label{fig:coordinates}
(Left figure): A unit sphere with a center O which is also the center of a black hole and a ray emitted at a point $\mathrm{E^\prime}$ which is at the inner edge of a disk.
 We introduce coordinates (X,Y,Z) so that the disk is on a plane OXY. We assume that an observer is at $\mathrm{O}^\prime$ with an angle $i \equiv \angle \mathrm{ZOO^\prime}$. 
 The domain of $i$ is $0 \leq i \leq \pi/2$.
 We also introduce ($\mathrm{X^\prime}, \mathrm{Y^\prime}$, $\mathrm{Z^\prime}$) so that a plane $\mathrm{O X^\prime Y^\prime}$ is overlap to a plane $\mathrm{OO^\prime E^\prime}$,
 where E is an intersection point of a line $\mathrm{OE^\prime}$ and the unit sphere.
 Angles $\alpha$, $\gamma$, $\psi$, and $\xi$ are defined as $\alpha \equiv \angle \mathrm{X^\prime OW}$, $\gamma \equiv \angle \mathrm{O^\prime OE^\prime}$, 
 $\psi \equiv \angle \mathrm{EOY}$, and $\xi \equiv \angle \mathrm{YEY^\prime}$, respectively, and where W is an intersection point of a line $\mathrm{Z^\prime O}$ and the unit sphere.
 The angle $\xi$ is defined on the unit sphere. 
 (Right figure): A unit celestial sphere with a center $\mathrm{O}^\prime$ and celestial coordinates $(\alpha, \beta)$ for the observer at $\mathrm{O^\prime}$.   
Coordinates $(x,y,z)$ are set so that $x$ and $y$ axes are parallel to lines $\mathrm{OX^\prime}$ and $\mathrm{OZ^\prime}$, respectively, and that $z$ axis overlaps with a line $\mathrm{OO^\prime}$.  
Note the celestial coordinate $\alpha$ is given by $\alpha=\angle \mathrm{X^\prime OW}$ and $\beta$ is defined for a domain $0 \leq \beta \leq \pi/2$ as an angle between the $z$ axis and the tangent line of the ray at $\mathrm{O^\prime}$.  
We define a point Q by an intersection point of the tangent line of the ray at $\mathrm{O^\prime}$ and the unit celestial sphere 
and we define a point P, which is identified by $(\alpha, \beta/2)$, on a $xy$ plane ($z=1$ plane).     
}
\end{figure*}
\section{Setup for imaging}
In the left panel of Fig.~\ref{fig:coordinates}, we set up coordinates for an image of a ray emitted by an inner edge of an optically and geometrically thin accretion disk at $\mathrm{E^\prime}$ seen 
by an observer at $\mathrm{O^\prime}$ with an angle~$i$ as well as Hioki and Miyamoto~\cite{Hioki:2022mdg} which is similar to Luminet~\cite{Luminet_1979}. 
We introduce a unit sphere with a center O which is also the centers of a black hole and coordinates (X,Y,Z).
We assume that the disk is on a plane OXY, its inner edge is the ISCO orbit, i.e., $r_\mathrm{e}=r_\mathrm{ISCO}$, 
and a distance between the observer $\mathrm{O^\prime}$ and the center O is $\mathrm{OO^\prime}=r_\mathrm{o}=10^5 M$. 
We choose the coordinate $Z$ to be $0 \leq i=\angle \mathrm{Z O O^\prime} \leq \pi/2$.
We have set the azimuthal angle coordinate $\varphi$ so that the azimuthal angle at $\mathrm{O^\prime}$ has the value of $\pi/2$, i.e., $\varphi_\mathrm{o}=\pi/2$. 
Let $\mathrm{E^\prime}$ be a point at the inner edge of the disk 
and E be the intersection point of a line $\mathrm{OE^\prime}$ and the unit sphere.
The azimuthal angle at $\mathrm{E^\prime}$ is given by $\varphi_\mathrm{e}= \pi/2-\gamma$, where $\gamma = \angle \mathrm{O^\prime OE^\prime}$ is the deflection angle.  
We introduce $(\mathrm{X^\prime}, \mathrm{Y^\prime}, \mathrm{Z^\prime})$ on the unit sphere so that a plane $\mathrm{O X^\prime Y^\prime}$ is equal to a plane $\mathrm{OO^\prime E^\prime}$. 
We define an angle $\xi \equiv \angle \mathrm{YEY^\prime}$ on the unit sphere and angles $\psi \equiv \angle \mathrm{EOY}$ and $\alpha \equiv \angle \mathrm{X^\prime OW}$, 
where W is the intersection of a line $\mathrm{Z^\prime O}$ and the unit sphere. 

We introduce a unit celestial sphere, coordinates $(x,y,z)$, and celestial coordinates $(\alpha, \beta)$ for the observer at $\mathrm{O^\prime}$ as shown in the right panel of Fig.~\ref{fig:coordinates}.   
Here, we set $(x,y,z)$ so that $x$ and $y$ axes are parallel to lines $\mathrm{OX^\prime}$ and $\mathrm{OZ^\prime}$, respectively, and that $z$ axis overlaps with a line $\mathrm{OO^\prime}$.  
The celestial coordinate $\beta$ is given by an angle between the $z$ axis and a line $\mathrm{O^\prime Q}$,
where Q is the intersection of the tangent line of the ray at $\mathrm{O^\prime}$ and the unit celestial sphere.
\begin{figure*}[t]
\includegraphics[width=0.32\linewidth]{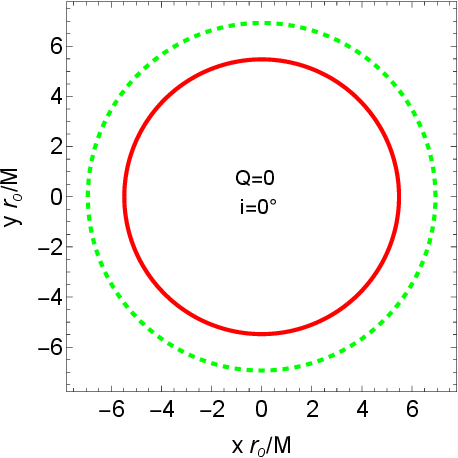}
\includegraphics[width=0.32\linewidth]{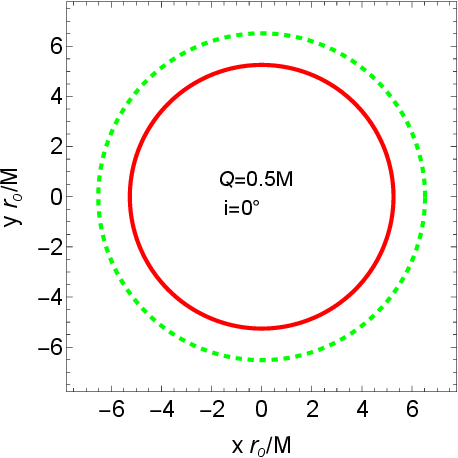}
\includegraphics[width=0.32\linewidth]{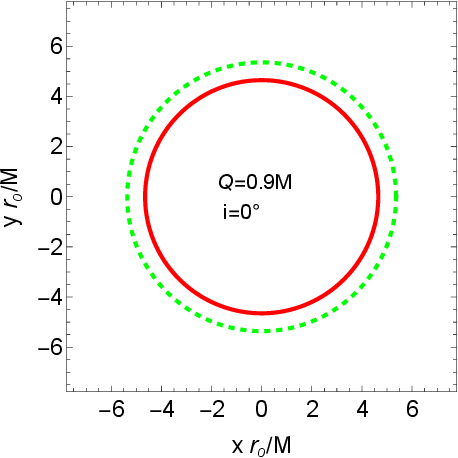}
\caption{
\label{fig:i=0}
Direct and lensing rings seen from $i=0^\circ$ are plotted by dashed~(green) and solid~(red) circles, respectively.
We set $r_\mathrm{e}=r_\mathrm{ISCO}$ and $r_\mathrm{o}=10^5M$. 
The black hole with a charge $Q=0$, $0.5M$, and $0.9M$, are shown in left, middle, and right panels, respectively. 
}
\end{figure*}

From the law of sine on the spherical triangles $\mathrm{EXX^\prime}$ and $\mathrm{EYY^\prime}$ we obtain, 
\begin{equation}\label{eq:sine1}
\frac{\sin \frac{\pi}{2}}{\sin \left( \frac{\pi}{2} -\psi \right)} = \frac{\sin \xi}{\sin \left( \frac{\pi}{2} -\alpha \right)}
\end{equation}
and 
\begin{equation}\label{eq:sine2}
\frac{\sin \xi}{\sin \left( \frac{\pi}{2} -i \right)} = \frac{\sin  \frac{\pi}{2}}{\sin \gamma},
\end{equation}
respectively, and then we eliminate $\xi$ from Eqs.~(\ref{eq:sine1}) and (\ref{eq:sine2}) to obtain 
\begin{equation}\label{eq:sine3}
\cos \alpha= \frac{\cos \psi \cos i}{\sin \gamma}.
\end{equation}
Let H be the foot of a perpendicular from $\mathrm{Y^\prime}$ to a line OE to obtain $\mathrm{OH}=\mathrm{OY^\prime} \cos \gamma =\cos \gamma$
and let K be the foot of a perpendicular from $\mathrm{Y^\prime}$ to a line OY to get $\mathrm{OK}=\mathrm{OY^\prime} \sin i =\sin i$ and $\mathrm{OH}=\mathrm{OK} \cos \psi =\sin i \cos \psi$.
Therefore, from the length of OH, we find
\begin{equation}\label{eq:cosgamma}
\cos \gamma= \sin i \cos \psi.
\end{equation}
From Eqs.~(\ref{eq:sine3}) and (\ref{eq:cosgamma}), we get 
\begin{equation}\label{eq:cosalpha}
\cos \alpha= \cot i \cot \gamma.
\end{equation}

The equation of the trajectory~(\ref{eq:trajectory1}) at the position of the observer $\mathrm{O^\prime}$, i.e., $r=r_\mathrm{o}=10^5 M$, becomes
\begin{equation}\label{eq:trajectory7}
\dot{t}_\mathrm{o}^2 =\frac{\dot{r}_\mathrm{o}^2}{f_\mathrm{o}^2  }+ \frac{r_\mathrm{o}^2  \dot{\varphi}_\mathrm{o}^2 }{f_\mathrm{o}},
\end{equation}
where $\dot{t}_\mathrm{o}$, $\dot{r}_\mathrm{o}$, $\dot{\varphi}_\mathrm{o}$, and $f_\mathrm{o}$ denote $\dot{t}$, $\dot{r}$, $\dot{\varphi}$, and $f(r)$ at $\mathrm{O^\prime}$, respectively.    
From the angle $\beta$ and Eq.~(\ref{eq:trajectory7}), 
$\dot{r}_\mathrm{o}$ and $\dot{\varphi}_\mathrm{o}$ can be written in
\begin{eqnarray}\label{eq:dotrbarphi}
&&\dot{r}_\mathrm{o}=\dot{t}_\mathrm{o} f_\mathrm{o} \cos \beta, \\
&&\dot{\varphi}_\mathrm{o}=\dot{t}_\mathrm{o} \frac{\sqrt{f_\mathrm{o}}}{r_\mathrm{o}} \sin \beta,
\end{eqnarray}
and, hence, we obtain 
\begin{eqnarray}\label{eq:tanbeta}
\tan \beta = \frac{\dot{\varphi}_\mathrm{o}}{\dot{r}_\mathrm{o}} r_\mathrm{o} \sqrt{f_\mathrm{o}}.
\end{eqnarray}
At $\mathrm{O^\prime}$, from Eqs.~(\ref{eq:trajectory4}) and (\ref{eq:effective}), we obtain
\begin{equation}\label{eq:dotro}
\dot{r}_\mathrm{o}=E \sqrt{1- f_\mathrm{o} \frac{b^2}{r_\mathrm{o}^2} },
\end{equation}
and, from Eq.~(\ref{eq:defL}), 
\begin{equation}\label{eq:dotvarphio}
\dot{\varphi}_\mathrm{o}=\frac{L}{r_\mathrm{o}^2}.
\end{equation}
From Eqs.~(\ref{eq:tanbeta})-(\ref{eq:dotvarphio}), we get 
\begin{eqnarray}\label{eq:tanbeta2}
\tan \beta = \frac{b\sqrt{f_\mathrm{o}}}{\sqrt{r_\mathrm{o}^2-f_\mathrm{o}b^2}}.
\end{eqnarray}

By the inscribed angle theorem on the angle $\beta$,
we project the point Q, which is determined by $(\alpha, \beta)$, onto a point P, which is specified $(x(\alpha,\beta), y(\alpha,\beta))$, on a plane $z=1$~\cite{Grenzebach:2014fha}.
From a simple calculation, we obtain $(x(\alpha,\beta), y(\alpha,\beta))$ as
\begin{eqnarray}
&&x=-2\sin \alpha \tan \frac{\beta}{2},\label{eq:x}\\
&&y=-2\cos \alpha \tan \frac{\beta}{2}\label{eq:y}.
\end{eqnarray}

Under the assumptions of $r_\mathrm{o} \gg M$ and $r_\mathrm{o} \gg \left| b \right| $, 
we get the relations,
\begin{eqnarray}
&&x\, r_\mathrm{o} \sim -b\sin \alpha,\label{eq:xro}\\
&&y\, r_\mathrm{o} \sim -b\cos \alpha.\label{eq:yro}
\end{eqnarray}

\section{Direct and lensing rings}
Due to the photon sphere, infinite numbers of rays emitted at a point on the disk can reach to the observer.
The rays cross the disk plane the $n$ times after the emissions from the disk.
We call ring images formed by the rays which cross the disk plane no time, once, and more than once as 
a direct ring~($n=0$), a lensing ring~($n=1$), and a photon ring~($n>1$), respectively.
The size of the photon rings is almost the same as the photon sphere and the former is equal to the latter in an infinite crossing limit $n \rightarrow \infty$ . 
Given the angle $i$,  
we can make the appearance of the every ring image 
by finding the impact parameter $b$ which satisfies Eq.~(\ref{eq:cosalpha}) with deflection angle $\gamma$ shown as Eq.~(\ref{eq:deflection_angle3}) or (\ref{eq:deflection_angle1}).

\begin{figure*}[t]
\includegraphics[width=0.32\linewidth]{Q05i00.eps}
\includegraphics[width=0.32\linewidth]{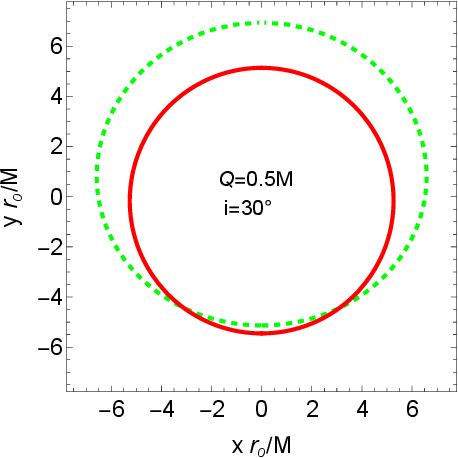}
\includegraphics[width=0.32\linewidth]{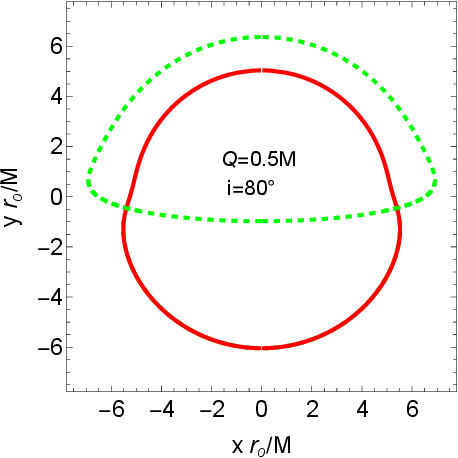}
\caption{
\label{fig:Q=0.5M}
The direct and lensing rings of the black hole with $Q=0.5M$ are denoted by dashed~(green) and solid~(red) circles, respectively.
We assume $r_\mathrm{e}=r_\mathrm{ISCO}$ and $r_\mathrm{o}=10^5M$. 
The rings with $i=0^\circ$, $30^\circ$, and $80^\circ$ are shown in left, middle, and right panels, respectively. 
}
\end{figure*}
\subsection{Rings seen from $i=0^\circ$}
First, we consider ring images seen from the angle~$i=0^\circ$.
In this case, the ring images have completely circular shapes as shown in Fig.~\ref{fig:i=0} 
and we confirm that their dimensionless radii are given by $\sim \left|b \right| /r_\mathrm{o}$.
Due to the lensing configuration of the black hole, the observer, and the disk,
the deflection angle of the observed ray is $\left| \gamma \right|= (1/2 +n)\pi$. 
Thus, the direct ring ($n=0$) and the lensing ring ($n=1$) have the deflection angles $\left| \gamma \right| = \pi/2$ and $3\pi/2$, respectively.
For each crossing number $n$, we can find the impact parameter $b$ to plot the ring image. 
We numerically solve Eq.~(\ref{eq:deflection_angle1}) for a given value of $Q$ to find 
the value of $b$ satisfying $|\gamma|=\pi/2$ which corresponds to the direct ring. 
In doing so, we find that
its absolute value $\left| b \right|$ for the direct image monotonically decreases from 
$6.93 M$ to $4.81 M$ as the charge $\left| Q \right|$ increases from $0$ to $M$.
The impact parameter for the lensing ring can be determined by numerically solving 
Eq.~(\ref{eq:deflection_angle3}) with $|\gamma|=3\pi/2$ for a given value of $Q$. 
We find that the value of $\left| b \right|$ monotonically decreases from $5.48M$ to 
$4.37M$ as the charge $\left| Q \right|$ increases from $0$ to $M$.

As a reference, we comment on a photon ring with $n \rightarrow \infty$
which corresponds to the image of the photon sphere at $r=r_\mathrm{ph}$ given in Eq.~(\ref{eq:rph}).
For the photon ring, $P=r_\mathrm{ph}$ should hold. Thus, we obtain $\left| b \right| =b_\mathrm{ph}$ 
where $b_\mathrm{ph}$ is given by Eq.~(\ref{eq:trajectory5}).
The impact parameter $\left| b \right|$ monotonically decreases from $3\sqrt{3}M \sim 5.196M$ to $4M$ 
as the charge $\left| Q \right|$ increases from $0$ to $M$.

\subsection{Rings seen from $i \neq 0^\circ$}
\begin{figure}[h]
\begin{center}
\includegraphics[width=\linewidth]{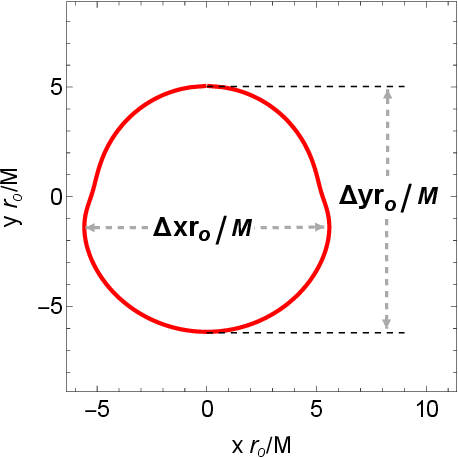}
\end{center}
\caption{
\label{fig:deftbar}
From the plot of the lensing ring, we read its width $\Delta x r_\mathrm{o}/M$ and its height $\Delta y r_\mathrm{o}/M$.}
\end{figure}

Second, we consider the direct ring ($n=0$) and the lensing ring ($n=1$) seen 
from the angle~$i \neq 0^\circ$ as shown in Fig.~\ref{fig:Q=0.5M}. 
If $i \sim 90^\circ$, the lensing ring has a shape that is close to a circle but not 
completely circular while the direct ring has a half circular one. 
Hereinafter, we concentrate on the lensing ring since we are interested in the constraint of the charge $Q$ by using the lensing ring.  
Given the fact that the lensing ring is not completely circular for the angle $i\neq0^\circ$, Eqs.~(\ref{eq:xro})-(\ref{eq:yro}) show that there is a range in the value of $b$ even within the same ring image. This is in contrast to the case for the angle $i=0^\circ$ where the ring image is exactly circular, and hence, $b$ can be uniquely determined for a given value of $Q$. 
Indeed, in drawing the lensing ring in Fig.~\ref{fig:Q=0.5M}, we numerically solve Eq.~(\ref{eq:deflection_angle3}) 
for a given value of $i$ to find the value of $b$ satisfying $\pi<|\gamma|\leq2\pi$. 
This process leads to a set of numbers $(\gamma,b)$ which determines $x$ and $y$ given in 
Eqs.~(\ref{eq:x})-(\ref{eq:y}), respectively, in the use of Eqs.~(\ref{eq:cosalpha}) 
and (\ref{eq:tanbeta2}). The direct ring can also be obtained in the similar way but by 
numerically solving Eq.~(\ref{eq:deflection_angle1}), instead of (\ref{eq:deflection_angle3}), 
to find the value of $b$ satisfying $0<|\gamma|\leq\pi$. Thus,
we define the average of the specific impact parameter $\left| \bar{b} \right|/M$ for the lensing ring by 
\begin{eqnarray}
\frac{\left| \bar{b} \right|}{M} \equiv \frac{\Delta x\, r_\mathrm{o} + \Delta y\, r_\mathrm{o}}{4M},
\end{eqnarray}
where $\Delta x\, r_\mathrm{o}/M$ and $\Delta y\, r_\mathrm{o}/M$ are the height and the width of the lensing ring, respectively, as shown in Fig.~\ref{fig:deftbar}.

\begin{figure}[t]
\begin{center}
\includegraphics[width=\linewidth]{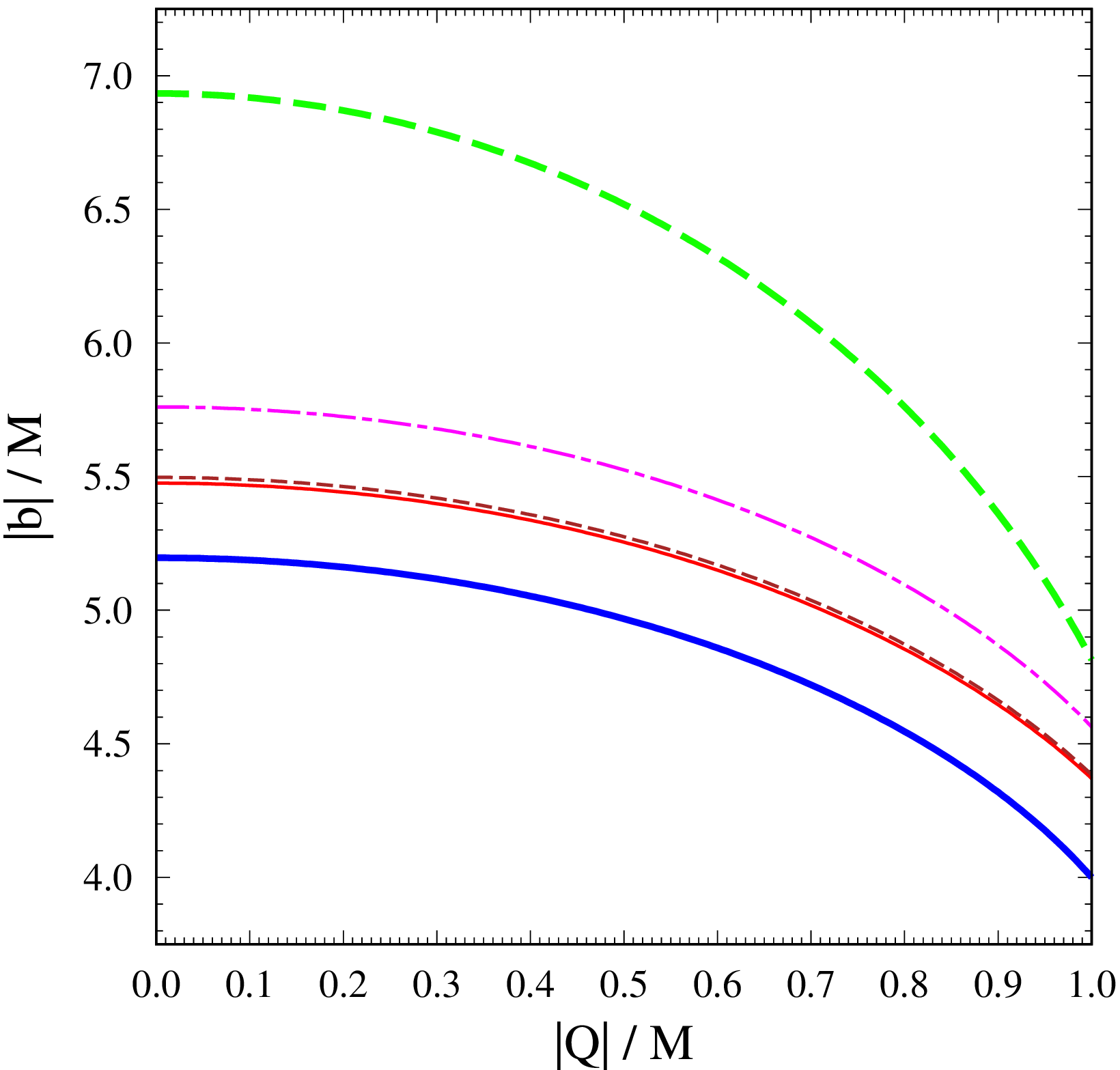}
\end{center}
\caption{
\label{fig:Qb}
Specific impact parameters $\left| b \right|/M$ against the specific charge $\left| Q \right|/M$ are shown.
Each curve corresponds to the direct ring with $i=0^\circ$ (thick-dashed green line), the lensing ring with $i=80^\circ$ (thin-chained magenta line), $i=30^\circ$ (thin-dashed brown line), $i=0^\circ$ (thin-solid red line), and the photon sphere (thick-solid blue line).
 See Sec.~IV~B for the specific impact parameters with $i=80^\circ$ and $30^\circ$.
}
\end{figure}

The averaged specific impact parameters $\left| \bar{b} \right|/M$ for the lensing ring with the angle $i=30^\circ$ and $80^\circ$ are plotted in Fig.~\ref{fig:Qb}. 
On this paper, the averaged specific impact parameters $\left| \bar{b} \right|/M$ are denoted by $\left| b \right|/M$ as a shorthand. 
We also plot the impact parameters for the direct ring and the photon sphere with the angle $i=0^\circ$. 
The latter is given by Eq.~(\ref{eq:trajectory5}) with (\ref{eq:rph}). 
Let us define the changing rate of impact parameter with respect to the charge $Q$ as 
\begin{equation}
\rho = \frac{b}{b_0}\,,\label{defrho}
\end{equation}
where $b_0$ is the value of $b$ in the absence of charge, i.e., $Q=0$. 
Figure~\ref{fig:Qb} shows that $\rho$ is the decreasing function with respect to $Q$ 
without relation to the angle $i$ nor number of times rays cross the disk plane. 
However, the behavior of $\rho$ around $Q/M\sim1$ differs between the lensing 
ring and photon sphere. Figure~\ref{fig:Qb} shows that the changing rate for the lensing rings 
with $i=0^\circ$, $30^\circ$, and $80^\circ$ is much milder than that for the photon sphere. 
As we will see in the next section, this property of the lensing ring can alleviate 
the relatively stringent constraints on the value of charge $Q$ obtained in 
Refs.~\cite{EventHorizonTelescope:2021dqv,EventHorizonTelescope:2022xqj}.

\section{Constraints of the charge by observations}
In this section, we summarize the observations on M87* and Sgr~A* used 
in Refs.~\cite{EventHorizonTelescope:2021dqv,EventHorizonTelescope:2022wkp,Vagnozzi:2022moj,daSilva:2023jxa}
and then, we give constrains on their charges by the change rate of the lensing ring.    

\subsection{M87*}
Gebhardt~\textit{et al.}~\cite{Gebhardt:2011yw} adopted a distance to M87* of $D= 17.9$ Mpc and estimated the mass of $M=(6.6 \pm 0.4) \times 10^9 M_\odot$, where $M_\odot$ is the solar mass, by stellar dynamics observations 
with laser adaptive optics to feed the Gemini telescope integral-field spectrograph and near-infrared integral field spectrograph
while Walsh~\textit{et al.}~\cite{Walsh:2013uua} estimated the mass of $M=(3.5^{+0.9}_{-0.7}) \times 10^9 M_\odot$ by gas-dynamical mass measurements from Hubble Space Telescope data acquired with the Space Telescope Imaging Spectrograph under the same assumption of the distance $D=17.9$ Mpc.

The EHT Collaboration~\cite{EventHorizonTelescope:2019dse,EventHorizonTelescope:2019ggy} observed the ring image of M87* with a diameter of $42 \pm 3$~$\mu$as at a wavelength of $1.3$ mm
and they concluded that the observed ring image is consistent with the Kerr black hole with $M=(6.5 \pm 0.7) \times 10^9 M_\odot$ in ray-traced GRMHD simulations
under the assumption of a distance $D=16.8 \pm 0.8$ Mpc in the $1\sigma$ level.~\footnote{
Very recently, the EHT Collaboration reported a ring image of M87* with a diameter of $43.3^{+1.5}_{-3.1} \mu$as at a wavelength of 1.3 mm~\cite{18January2024}.  
We do not use the observations to constrain the charge on this paper since they do not perform a new mass estimation which needs a dedicated calibration using GRMHD simulations.
}
In Ref.~\cite{EventHorizonTelescope:2019pgp}, they 
confirmed that the observed asymmetric ring is consistent with the gravitational lensing of synchrotron radiations from a hot plasma near the black holes as had been expected~\cite{Yuan:2014gma}.
In Ref.~\cite{EventHorizonTelescope:2019ggy}, 
they obtained a ratio of the mass $M$ to the distance $D$ as $\theta_\mathrm{g} \equiv M/D =3.8\pm 0.4$~$\mu$as
from the EHT observations, crescent model fitting, GRMHD model fitting, and image domain feature extraction.
They calculated a ratio $\theta \equiv M/D$ of $3.62^{+0.60}_{-0.34}$~$\mu$as from stellar dynamics observations~\cite{Gebhardt:2011yw} 
and $\theta$ of $2.05^{+0.48}_{-0.16}$~$\mu$as from gas dynamics observations~\cite{Walsh:2013uua}.
They compared between the ratios by
\begin{equation}
\delta \equiv \frac{\theta_\mathrm{g}}{\theta}-1,
\end{equation}
and they got the values of $\delta=-0.01\pm 0.17$ and $0.78\pm 0.3$ from the stellar and gas dynamics mass measurements, respectively, 
and they pointed out that the EHT observation is consistent with the stellar dynamics observation 
but it is not with the gas dynamics. 
In Ref.~\cite{EventHorizonTelescope:2021dqv}, the EHT Collaboration assumed that $\delta=0.00 \pm 0.17$ corresponds 
with the permissible difference of the size of the photon sphere of the Reissner-Nordstr\"{o}m black hole 
from the Schwarzschild black hole to get the constraint on the charge $\left|Q \right|/M<0.90$ in the $1\sigma$ level. 

\begin{figure}[htbp]
\includegraphics[width=\linewidth]{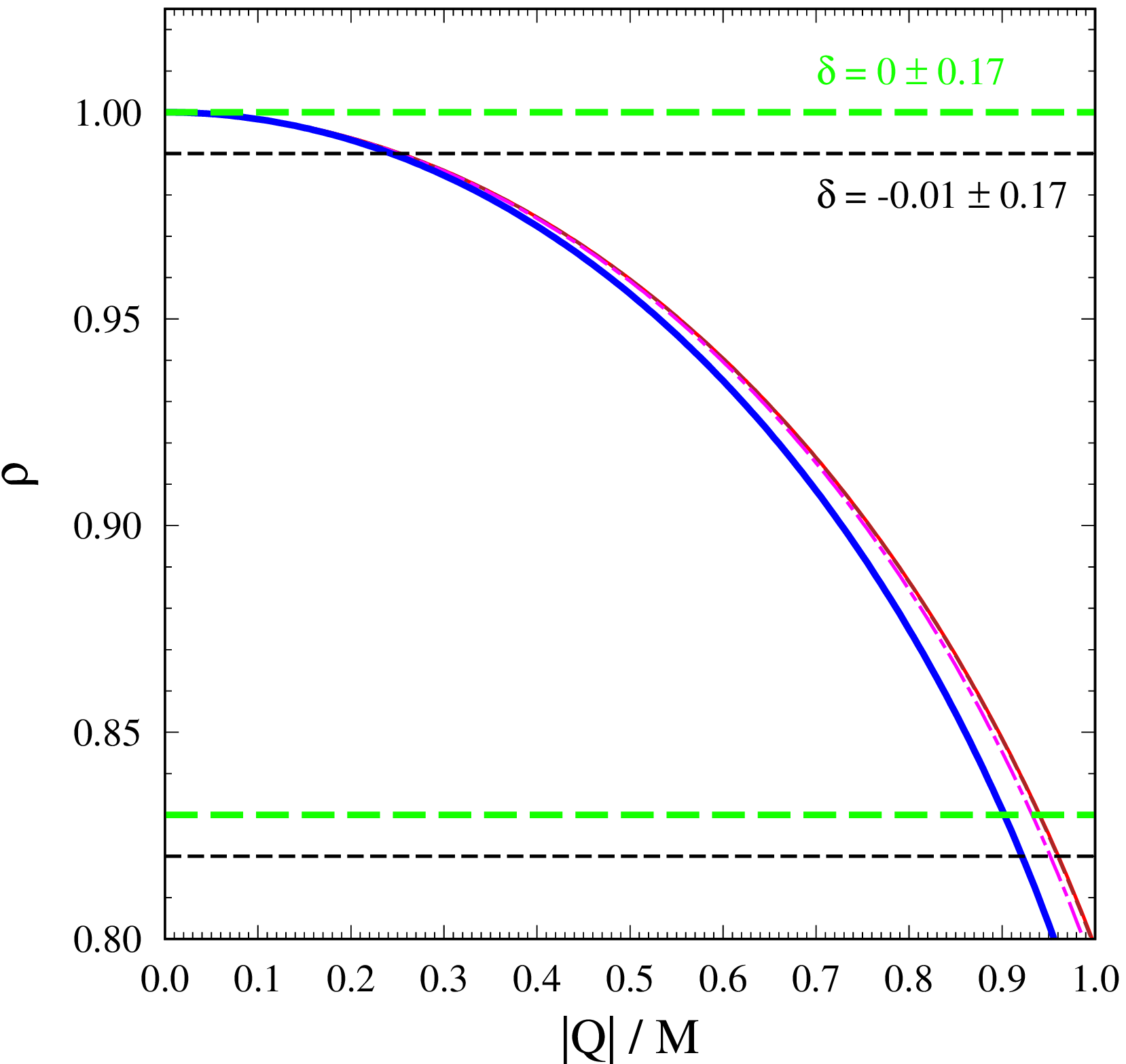}
\caption{
\label{fig:Qchangingrate}
The constraint of the charge by the observations of M87* in the $1\sigma$ level with $\delta=-0.01\pm 0.17$ which is denoted by thin-dashed (black) lines.
As a reference, $\delta=0.00 \pm 0.17$ used in Ref.~\cite{EventHorizonTelescope:2021dqv} is also denoted by thick-dashed (green) lines.
The curves represent the changing rate of the specific impact parameters, i.e., $\rho$ defined in Eq.~(\ref{defrho}), 
 with respect to the charge $Q$. Each curve corresponds to the lensing ring with $i=80^\circ$ (thin-chained magenta line), $i=30^\circ$ (thin-dashed brown line), $i=0^\circ$ (thin-solid red line), and the photon sphere (thick-solid blue line).
Notice that the curves for lensing rings with $i=80^\circ$, $30^\circ$, and $0^\circ$ are overlapped.
We find the bound of the charge $\left|Q \right|/M \leq 0.96$ by the lensing rings.   
We recover the constraint $\left|Q \right|/M \leq 0.90$ by the photon sphere in Ref.~\cite{EventHorizonTelescope:2021dqv}.}
\end{figure}

On the other hand, by using $\delta=-0.01\pm 0.17$ and the change rate of the radius of the lensing ring, we get a bound $\left|Q \right|/M \leq 0.96$ in the $1\sigma$ level as shown in Fig.~\ref{fig:Qchangingrate}.
As a reference, in Fig.~\ref{fig:Qchangingrate}, we also show the charge bound $\left|Q \right|/M<0.90$ by the EHT Collaboration~\cite{EventHorizonTelescope:2021dqv}.  

\subsection{Sgr~A*}
The distance and mass of Sgr~A* were estimated by $D=8277 \pm 9 \pm 33$~pc and $M=(4.297 \pm 0.013) \times 10^6 M_\odot$  by Very Large Telescope Interferometer (VLTI) observation~\cite{GRAVITY:202101,GRAVITY:2021xju} 
and $D=7935 \pm 50 \pm 32$~pc and $M=(3.951 \pm 0.047) \times 10^6 M_\odot$ by the Keck observation~\cite{Do:2019txf}.
The EHT Collaboration~\cite{EventHorizonTelescope:2022wkp} reported a ring image with the diameter of $51.8 \pm 2.3~\mu$as observed at the wavelength of $1.3$~mm in the $1\sigma$ level, 
and they estimated the mass of $M=4.0^{+1.1}_{-0.6} \times 10^6 M_\odot$ 
under an assumption of the discance $D=8.15 \pm 0.15$ kpc~\cite{EventHorizonTelescope:2022exc}
and the deviations of the EHT observation from the VLTI observations as $\delta=-0.08 \pm0.09$ and from the Keck observations as $\delta=-0.04^{+0.09}_{-0.10}$.
The EHT Collaboration~\cite{EventHorizonTelescope:2022xqj}
obtained a bound $0 \leq \left|Q \right|/M \leq 0.84$ from the Keck observations, which is stricter than the one from the VLTI observations,
by using the change rate of the radius of the photon sphere of the Reissner-Nordstr\"{o}m black hole in the same way as Ref.~\cite{EventHorizonTelescope:2021dqv}.
\begin{figure*}[htbp]
\includegraphics[width=0.49\linewidth]{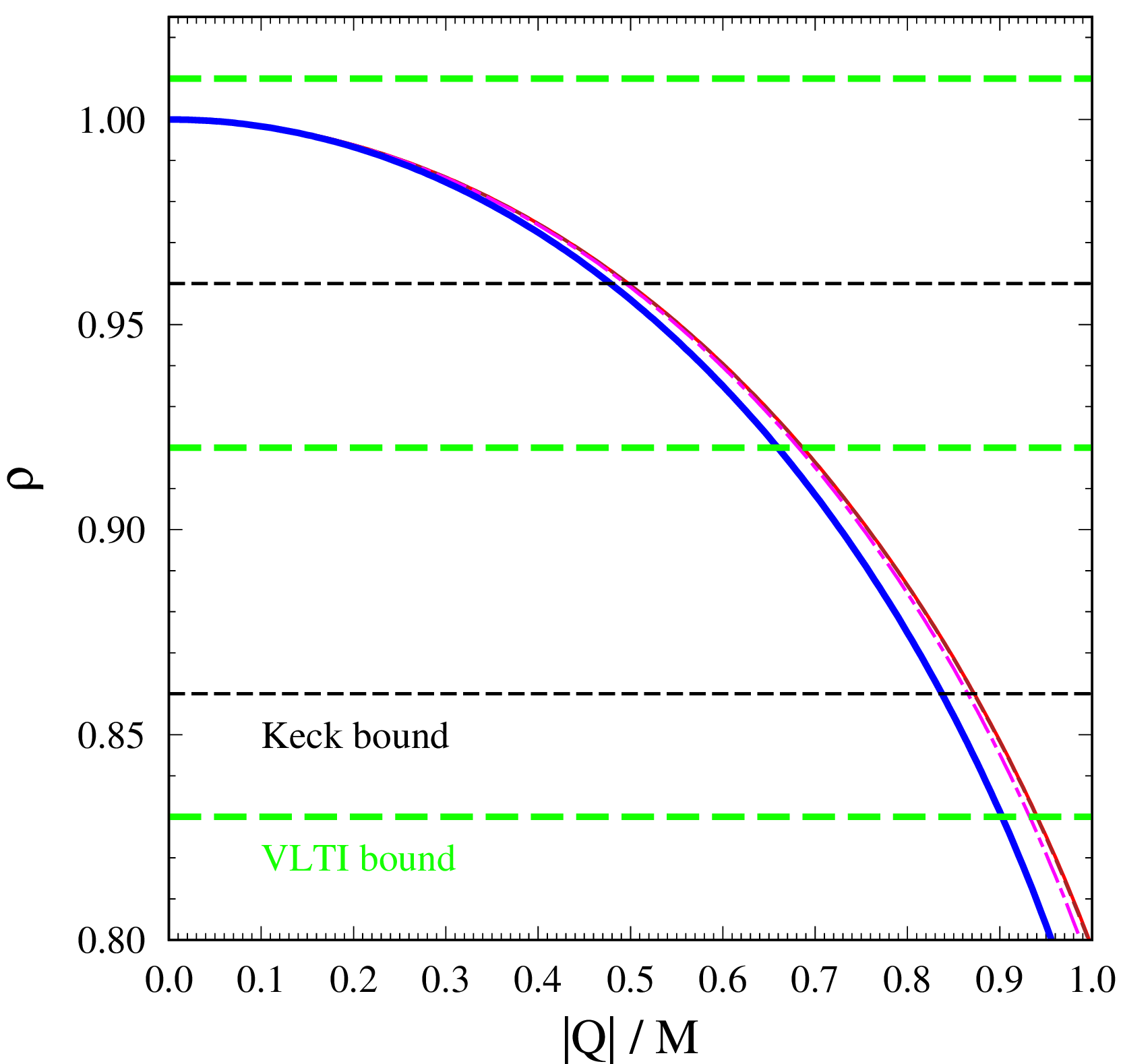}
\includegraphics[width=0.49\linewidth]{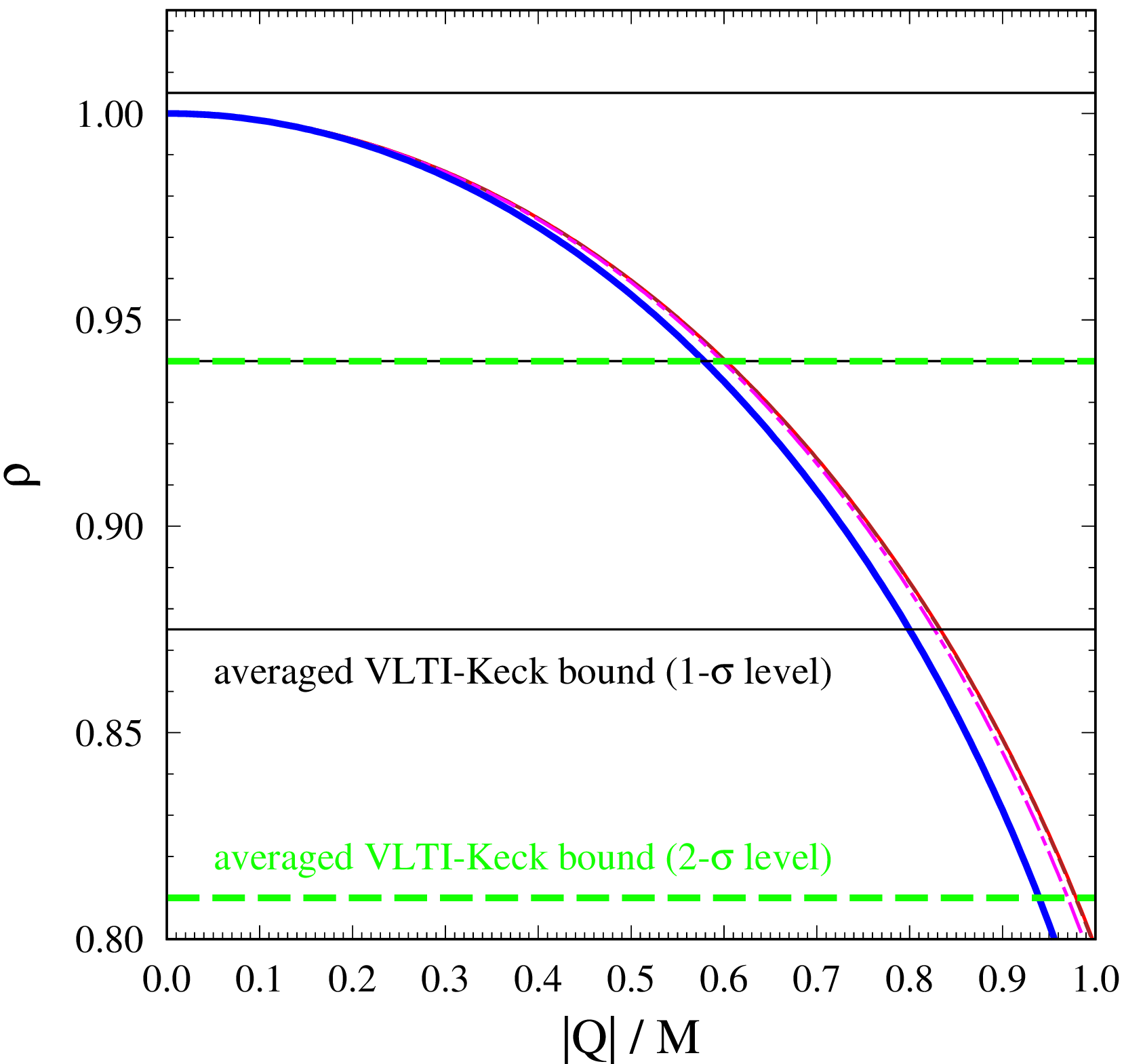}
\caption{
The constraints of the charge by the observations of Sgr~A*. 
Thin-chained~(magenta), thin-dashed~(brown), thin-solid~(red), and thick-solid~(blue) curves denote 
the change rate of the specific impact parameters $\left| b \right|/M$ for the lensing rings with $i=80^\circ$, $30^\circ$, and $0^\circ$, which are overlapped, and 
the photon sphere, respectively.
\textit{Left pannel}:
Deviations from the Keck observations $\delta=-0.04^{+0.09}_{-0.10}$ and from the VLTI observations $\delta=-0.08\pm0.09$ in $1$-$\sigma$ level 
are denoted by the thin-dashed~(black) and thick-dashed~(green) lines, respectively.
From Keck observations, we find the constraint $0 \leq \left| Q \right|/M \leq 0.87$ by the lensing rings and  
we recover the constraint $\left|Q \right|/M \leq 0.84$ in Ref.~\cite{EventHorizonTelescope:2022xqj} by the photon sphere.
\textit{Right pannel}:
The average values of $\delta$ from VLTI and Keck observations $\delta=-0.06 \pm 0.065$ in $1$-$\sigma$ level and $\delta=-0.06 \pm 0.13$ 
in $2$-$\sigma$ level~\cite{Vagnozzi:2022moj} are denoted by the thick-solid~(black) and the thick-dashed~(green) lines, respectively. 
We get the averaged VLTI-Keck bounds on the charge $0 \leq \left|Q \right|/M \leq 0.83$ in $1$-$\sigma$ level and $0 \leq \left|Q \right|/M \leq 0.97$ in $2$-$\sigma$ level by the lensing rings and   
we recover $0 \leq \left|Q \right|/M \leq 0.8$ in $1$-$\sigma$ level~\cite{Vagnozzi:2022moj} and $0 \leq \left|Q \right|/M \leq 0.939$ in $2$-$\sigma$ level~\cite{daSilva:2023jxa} by the photon sphere.
}
\end{figure*}
Vagnozzi~\textit{et al.} considered the average values of $\delta$ from VLTI and Keck observations as $\delta=-0.06 \pm 0.065$ in the $1\sigma$ level and $\delta=-0.06 \pm 0.13$ in the $2\sigma$ level, 
to get the constraints bound of the charge $0 \leq \left|Q \right|/M \leq 0.8$ in the $1\sigma$ level and  $0 \leq \left|Q \right|/M \leq 0.95$ in the $2\sigma$ level~\cite{Vagnozzi:2022moj}. 
The bound in $2$-$\sigma$ level
was recalculated as $0 \leq \left|Q \right|/M \leq 0.939$ by Da Silva~\textit{et al.}~\cite{daSilva:2023jxa}. 

From the change rate of the lensing ring, 
we get a Keck bound $0 \leq \left|Q \right|/M \leq 0.87$ 
and a VLTI bound $0 \leq \left|Q \right|/M \leq 0.93$ in the $1\sigma$ level and     
and averaged VLTI-Keck bounds $0 \leq \left|Q \right|/M \leq 0.83$ and $0 \leq \left|Q \right|/M \leq 0.97$ in the $1\sigma$ and $2\sigma$ levels, respectively, 
shown in Fig.~8.  
From Fig.~8, we also confirm the constraints by the change rate of the photon sphere 
in Refs.~\cite{EventHorizonTelescope:2022xqj,Vagnozzi:2022moj,daSilva:2023jxa}.

\section{Conclusion and Discussion} 
In this paper, we have investigated how to put bounds to the electrical or alternative charges of the black holes by the EHT observations and the other observations.
One may consider that GRMHD simulations should be used to get reliable constraints on the charges but the simulations need dedicated calibrations 
and it is not easy to perform in every spacetime and gravitational theory. 

The EHT Collaboration gave the constraints on their electrical or alternative charges of M87* and Sgr~A*, 
by using the change rate of the radius of the photon sphere~\cite{EventHorizonTelescope:2021dqv,EventHorizonTelescope:2022xqj,Vagnozzi:2022moj}. 
This method is simple and straightforward but 
we should keep in mind a fact that the observed ring is consistent with neither the photon sphere nor the photon ring. It is consistent with strong lensing images formed by synchrotron radiations from a hot plasma near the black holes according to their GRMHD simulations~\cite{EventHorizonTelescope:2019dse,EventHorizonTelescope:2019ggy}.
Their constraints by the change rate of the radius of the photon sphere would be more severe than 
other estimations by using the change rate of a radius which is larger than the photon sphere since the effect of the charge on observables would get weaker as the radius is larger.

We have showed that the constraint of the charges by the change rate of the photon sphere obtained in Refs.~\cite{EventHorizonTelescope:2021dqv,EventHorizonTelescope:2022xqj,Vagnozzi:2022moj} can be relaxed by comparing the change rate of size of the lensing ring formed by emissions from ISCO particles at the inner edge of geometrically thin disks around the black holes.
We admit that the geometrically thin disk with the inner edge at the ISCO of the particles is too simple 
to explain the asymmetry of observed ring images~\cite{EventHorizonTelescope:2019pgp} and 
their polarization~\cite{EventHorizonTelescope:2021bee,EventHorizonTelescope:2021srq,EventHorizonTelescope:2023fox,EHT2024a,EHT2024b}
and we do not claim that the observed ring images are explained by the lensing rings. 
Our point is that the black hole and disk system will be enough useful to check 
the constraint on the charges estimated by the photon sphere~\cite{EventHorizonTelescope:2021dqv,EventHorizonTelescope:2022xqj,Vagnozzi:2022moj} and to show that the constraint can be relaxed.   

We have investigated the bound on the charge by the change rate of lensing ring formed by emissions from the ISCO in the Reissner-Nordstr\"{o}m black hole spacetime.
The inner edge of an accretion disk around black hole may not match the ISCO~\cite{Abramowicz2010,Wielgus:2021peu},
but a number of observational aspects of the ring images may be explained by the properties of accretion flow across the ISCO~\cite{Mummery:2024rtw}.
We do not discuss the effect of the ISCO on the ring images in the accretion flow since it is beyond the scope of this paper.    
Even if light sources with the largest contribution to the observed ring are not on the ISCO, 
if the reflectional points of their rays are outside the photon sphere, 
then, the bounds of the charge would be relaxed~\cite{EventHorizonTelescope:2021dqv,EventHorizonTelescope:2022xqj,Vagnozzi:2022moj}. 
This is because the effects of the electrical or alternative charge decreases as the reflectional points of the rays is far from the black hole.

\section*{Acknowledgements}
The authors are deeply grateful to Yuta Suzuka for useful discussion.
R.~K. is supported by the Grant-in-Aid for Early-Career Scientists 
of the JSPS No.~20K14471 and the Grant-in-Aid for Scientific 
Research (C) of the JSPS No.~23K034210. 
\appendix
\section{$r_\mathrm{ISCO}$ and $P$}
The analytical forms of $r_\mathrm{ISCO}$ and $P$ can be expressed as
\begin{eqnarray}
r_\mathrm{ISCO}
=2M+p_0+\frac{4M^2-3Q^2}{p_0}
\end{eqnarray}
and 
\begin{eqnarray}
P=\frac{1}{2} \sqrt{\frac{2 b^2}{3}+p_2} 
+\frac{1}{2} \sqrt{\frac{4 b^2}{3}-p_2-\frac{4 b^2 M}{\sqrt{\frac{2 b^2}{3}+p_2}}},
\end{eqnarray}
respectively, and where $p_0$, $p_1$, $p_2$, and $p_3$ are given by
\begin{eqnarray}
p_0&\equiv& \sqrt[3]{ \frac{ 2 Q^4-9 Q^2 M^2 +Q^2 p_1 +8M^4 }{M} }, \\
p_1&\equiv& \sqrt{ 4 Q^4-9 Q^2 M^2+5M^4 }, \\ 
p_2&\equiv& \frac{b \sqrt[3]{- b^3+54 b M^2-36 b Q^2-p_3}}{3} \nonumber\\
&&+\frac{ b \left(b^2-12 Q^2\right)}{3 \sqrt[3]{-b^3+54bM^2-36 bQ^2-p_3}}, \\
p_3 &\equiv& \sqrt{ b^2 \left(b^2-54 M^2+36 Q^2\right)^2- \left(b^2-12 Q^2\right)^3}. \nonumber\\
&&
\end{eqnarray}
If the charge vanishes $Q=0$, we get $r_\mathrm{ISCO}=6M$ and
\begin{eqnarray}
P=\frac{\sqrt[3]{-3} b^2-(-1)^{2/3} \left(9 b^2 M+\sqrt{81 b^4 M^2-3 b^6}\right)^{2/3}}{3^{2/3} \sqrt[3]{9 b^2 M+\sqrt{81 b^4 M^2-3 b^6}}}. \nonumber\\
\end{eqnarray}

\section{A ring image of M87* at wavelength of $3.5$~mm}
We comment on a ring image of M87* by very-long-baseline interferometry~(VLBI) observations with the Global Millimetre VLBI Array complemented by the phased Atacama Large Millimetre/submillimetre Array and the Greenland Telescope at wavelength of $3.5$~mm reported in Ref.~\cite{Lu:2023bbn}. The ring has a diameter of $64^{+4}_{-8}$~$\mu$as and  
they assumed a distance of $D=16.8$~Mpc and a mass of $M=6.5 \times 10^9 M_\odot$. 
They claimed that it is natural to assume that the black hole is at the center of the ring. 
They assumed that the nonthermal synchrotron model from the jet and the thermal synchrotron model 
from the accretion flow. In the both models, plasma surround the black hole 
is optically thin at $1.3$~mm and optically thick at $3.5$~mm due to the synchrotron self-absorption of the plasma. 
They found that the the thermal model can match the observed ring size but the nonthermal model makes smaller 
($\geq 30\%$) than the observed ring. 
We note that the observed ring size can be read as $\left| b \right|/M=8.33^{+0.52}_{-1.04}$ and it is larger than 
the impact parameter of the direct ring of ISCO seen from the angle $i=0^\circ$ as shown Fig.~9.  
Thus, we realize that the observed ring size is too large to constrain the charge of the Reissner-Nordstr\"{o}m black hole. 
\begin{figure}[htbp]
\includegraphics[width=\linewidth]{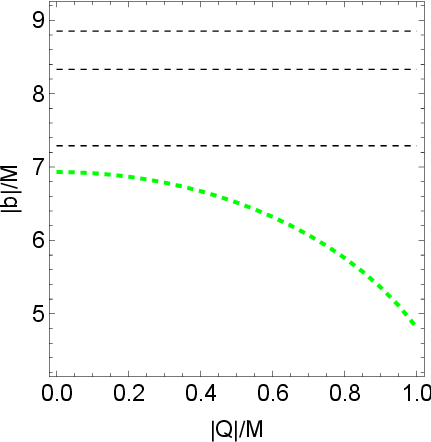}
\caption{
Comparison sizes between the observed ring of M87* at $3.5$~mm and the impact parameter of direct ring of ISCO seen the angle $i=0^\circ$.  
A thick-dashed~(green) curve and thin-dashed~(black) lines denote the specific impact parameters $\left| b \right|/M$ of the direct ring  
and of the observed ring image~$\left| b \right|/M=8.33^{+0.52}_{-1.04}$, respectively.}
\end{figure}

\end{document}